\documentclass[aps,pra,twocolumn,groupedaddress,amsmath,showpacs]{revtex4}
\usepackage{graphicx}

\begin{document}


\title{Stark-modulation spectroscopy of the B(1)$\left[^{3}\Pi\right]$
 state of PbO}



\author{D. Kawall,\footnote{Present Address : Department of Physics, University of Massachusetts,
Amherst, MA 01003 and RIKEN-BNL Research Center, Upton, NY 11973}
Y. V. Gurevich, C. Cheung, S. Bickman, Y. Jiang, and D. DeMille}
%
\affiliation{Department of Physics, Yale University, P.O. Box
208120, New Haven, CT 06520-8120}
%

\date{\today}

\begin{abstract}
We report detailed spectroscopic measurements of the $X(0)
\left[^1\Sigma^+\right](v=0) \rightarrow
B(1)\left[^{3}\Pi\right](v=5)$ transition in PbO. Using a
Stark-modulated laser absorption technique, we have measured the
hyperfine constant of $^{207}$PbO in the $B(1)$ state, as well as
the $B(1)(v=5)$ rotational constant, $X-B$ isotope shifts, etc.
The hyperfine constant of the $B(1)$ state is of interest as a
benchmark for calculations of PbO electronic structure, related to
experiments to search for the electric dipole moment of the
electron.
\end{abstract}

\pacs{33.15.-e,33.20.Kf,14.60.Cd}

\maketitle

%
%

Spectroscopy of the PbO molecule has become of interest since it
is a good candidate for use in searches for a permanent electric
dipole moment (EDM) of the electron, $d_{e}$. EDMs are interesting
because a non-zero value for an EDM of any fundamental particle
would violate parity and time-reversal symmetries. Standard model
predictions for EDMs are well below any proposed experimental
sensitivities. However, most extensions of the standard model
predict dramatically enhanced EDMs.

A promising approach towards improving the limits on $d_{e}$
involves heavy polar diatomic molecules in a configuration with an
unpaired electron spin. In such  systems, the effective electric
field $W_{d}$ seen by the unpaired electron can be many orders of
magnitude larger than external fields attainable in the laboratory
\cite{misha,titov}.  An electron EDM can be detected by spin
polarizing an electron along this internal field and searching for
the characteristic linear Stark shift $\Delta E=-d_{e} W_{d}$.
Interpreting experimental limits on $\Delta E$ measured in a
molecule in terms of $d_{e}$ requires a value for $W_{d}$. This
can be obtained using semi-empirical wavefunctions for the state
of interest \cite{misha}, or from {\it{ab initio}} calculations
\cite{titov}. Spectroscopic properties sensitive to the electron
spin density at the nucleus such as hyperfine structure (hfs) can
constrain the parameters in semi-empirical evaluations of $W_{d}$,
or test the predictions of {\it{ab initio}} calculations.

An experiment is underway in our lab using the
$a(1)\left[^{3}\Sigma^{+}\right]$ state of PbO, that may improve
the current limit $|d_{e}| <1.6\times 10^{-27}~e~$cm \cite{regan}
by several orders of magnitude \cite{daved}. Such an improvement
would probe large regions of the parameter space of many standard
model extensions. Stimulated by this work, a measurement of the
hyperfine constant of the $a(1)$ state was recently reported
\cite{larry}. The hfs of the $B(1)\left[^{3}\Pi\right]$ state of
$^{207}$PbO is also of interest, since it can be used to further
check and/or refine the electronic structure calculations used to
calculate the value of $W_d$ in the $a(1)$ state.  In addition,
knowledge of the $B(1)$ hfs can be used for estimating $W_{d}$ for
the $B(1)$ state, which also may be a viable candidate for an EDM
search \cite{egorov}. This motivates this effort to determine the
hfs of the $B(1)$ state of PbO.

The measurements were made by observing the absorption of laser
light by PbO vapor in a cell as the laser wavelength was tuned.
The cell contains PbO of natural isotopic abundance; it is
constructed from a 5.7 cm cube of alumina, with 5.1 cm diameter
through holes bored perpendicular to, and centered on, each face.
The top and bottom holes of the resulting cubical frame structure
are capped with alumina plates. Thin YAG windows are
bonded on the remaining four holes using gold foil as an
intermediate layer. The cell is heated to $700^{\circ}$C by the
radiation from resistively heated thin foils of tantalum supported
on a quartz framework surrounding the cell. This results in a PbO
number density $n\approx 4\times 10^{13}$ cm$^{-3}$ in the cell.
The laser used in the measurements was an external cavity diode
system based on a standard design \cite{malcolm}. Our system used
a Nichia NLHV500C violet diode laser and 3600 grooves/mm
diffraction grating in a Littrow configuration. The laser provided
$\lesssim$1 mW of power around 406.5 nm which passed through the
YAG windows into the cell.

To enable sensitive lock-in detection of the absorption, we
employed a  Stark modulation technique \cite{watanabe}.  As
explained below, this method enhances signals from low J levels
where the hfs is largest, which are otherwise difficult to isolate
within the spectrally congested region near the bandhead.  Here we
briefly discuss the basic features of the Stark-modulated signals,
as relevant to our experiment.

\begin{figure}
\includegraphics[width=7cm]{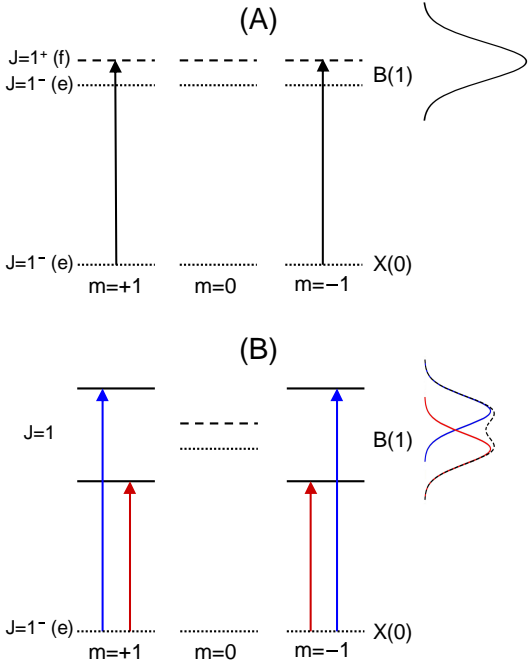}
{\caption{\label{fig:starkmod}Stark modulation of the excited
state energy levels, leading to a modulation of the transmitted
light intensity. Plot (A) shows the $Q1$ transition and the
absorption expected at $E$=0. In (B), with $E$ large, the Stark
effect mixes and shifts the doublet levels. The difference in
absorption is recorded in the scans.}}
\end{figure}

The Stark effect in the $X(0) \rightarrow B(1)$ transition is
dominated by shifts due to mixing between the closely-spaced
$\Omega$-doublet levels in the $B(1)$ state. The doublet members,
$e$ and $f$, share a common value of total (electronic +
rotational) angular momentum $\mathbf{J}$ but are of opposite
parity.  The spacing between the doublet levels is given by
$\Delta_{\Omega}(J)=qJ(J+1)$, where $q\ll B_{v=5}$, the rotational
constant for the $B(v=5)$ state.  [By convention, $q$ is positive
if the $e$ level, with parity $P=(-1)^J$, is at higher energy.] In
the absence of an electric field $\mathbf{E}$, selection rules for
the electric-dipole $X \rightarrow B$ transition ensure that the
laser light is absorbed only in the transition to one of the
doublet levels. In the presence of $\mathbf{E}$, the $e$ and $f$
levels mix and are shifted by an amount $\Delta\nu$ given by:
\begin{eqnarray}
\Delta\nu&=& \pm\left[
\sqrt{\left(\Delta_{\Omega}(J)/2\right)^{2}+(\alpha E)^{2}}
-\Delta_{\Omega}(J)/2\right]. \label{eqn:repulsion}
\end{eqnarray}
Here the upper (lower) sign corresponds to the upper (lower) level
in the limit $E\rightarrow 0$, and
\begin{eqnarray}
\alpha&=&\mu_{B(1)} M_{F}\frac{F(F+1)+J(J+1)-I(I+1)}
{2F(F+1)J(J+1)}, \label{eqn:alphadef}
\end{eqnarray}
where $\mu_{B(1)}$ is the dipole moment of the $B$ state, and
$F=J+I$ is the total angular momentum. In the limit of large $E$
and small $J$, $\Delta\nu\propto E$ and the doublet levels are
completely mixed. In this limit, the laser is absorbed with equal
strength by both eigenstates of the mixed doublet, and the
repulsion of the doublet levels results in a characteristic
broadening of the absorption spectral line (see Fig.
\ref{fig:starkmod}).
For a laser tuned to one of the unperturbed doublet levels, by
modulating $\mathbf{E}$, maxima in absorption will occur twice per
period at the minima of $E$. The $\Omega$-doublet mixing leads to
small shifts in the line centers, as does the slight residual
Stark mixing with nearby rotational levels of opposite parity.
Both of these small shifts ($\leq$ 50 MHz) are accounted for in
the analysis.

In our measurements, an electric field $\mathbf{E}$ was produced
using two parallel 3.8 cm diameter gold foil electrodes spaced 0.6
cm apart in the cell.  The laser light was linearly polarized
parallel to $\mathbf{E}$ and traversed the vapor cell between the
electrodes. A reference signal at $\omega=2\pi\times 99$ kHz was
amplified to produce a voltage $V=V_{0}\cos(\omega t)$, where
$V_{0}\approx 800$ V, across the electrodes. The transmitted light
was detected with a photodiode and amplified.  Part of the laser
beam was diverted for a wavelength measurement (using a Burleigh
WA-1500 Wavemeter), and the laser power reflected from the cell
was monitored with a second "reference" photodiode.

The transmitted signals were sampled at 1 MHz with a 12 bit ADC
and recorded for 4 s. The amplitude of the signal at the 2nd
harmonic of the reference was extracted from the digitized data.
Then the laser frequency was advanced roughly 150 MHz
by simultaneous adjustment
of the grating angle and diode current, and a new absorption
measurement was taken. This cycle was repeated until a mode-hop
occurred, yielding scans covering 15-30 GHz.

\begin{figure*}
\includegraphics[width=0.95\textwidth]{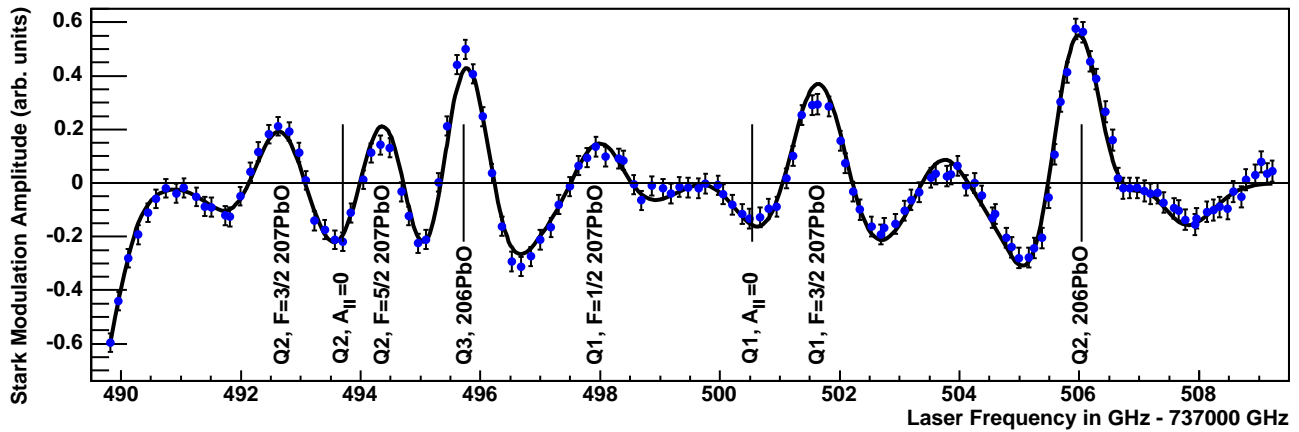}
{\caption{\label{fig:scan1} Segment of a typical Stark modulation scan. The
points are data and the line is a fit. The vertical bars mark the
line centers in the absence of hfs; the hyperfine splitting of the
$^{207}$PbO lines is clearly visible.}}
\end{figure*}

The scans showed absorption features corresponding to $^{208}$PbO,
$^{207}$PbO, and $^{206}$PbO molecules making $P$-, $Q$-, and
$R$-branch transitions from the electronic ground state
$X(0)\left[^{1}\Sigma^{+}\right](v''=0,J'')$ to the excited state
$B(1)\left[^{3}\Pi\right](v'=5,J')$, with $S/N \gtrsim 1$ for
$J''=0-6$.  An example is shown in Fig. {\ref{fig:scan1}}. Before
analyzing the data, the absorption amplitudes were corrected for
the variation in laser power with frequency during a scan. We
expect residual variations in amplitude over the scan of $\sim
5\%$, based on observed fluctuations in the ratio between
transmitted and reflected power (presumably due to interference
effects in the cell windows). The statistical uncertainty in each
measurement was estimated from the fluctuations of the second
harmonic out of phase with the signal, and was typically a few
times greater than the shot noise in the photodiode signal. The
largest signal was observed from the Q1 $^{208}$PbO line, with a
peak $S/N\ge$75, corresponding to $\approx$ 4 ppm absorption.
These scans were fit to extract the hfs constant and coefficients
in the Dunham expansion, as described below.

Hyperfine structure in the $B(1)$ state of $^{207}$Pb0 arises from
the interaction $H_{\mathrm{hfs}}=\mathbf{J}_{\mathrm{e}}\cdot
\mathbf{A}\cdot\mathbf{I} = A_{\parallel}J_{\mathrm{e}Z}I_{Z} +
A_{\perp}\left(J_{\mathrm{e}X}I_{X} +
J_{\mathrm{e}Y}I_{Y}\right)$, where $\mathbf{A}$ is the hyperfine
tensor, $I=1/2$ is the nuclear spin of $^{207}$Pb,
$\mathbf{J}_{\mathrm{e}}$ is the electronic angular momentum, and
the coordinate axes are defined in the body-fixed frame, with
$\hat{Z}$ along the internuclear axis $\hat{n}$. For levels with
low $J$ values in an $\Omega =1$ electronic state, the hfs is
dominated by the term in $H_{\mathrm{hfs}}$ proportional to
$A_{\parallel}$. The terms containing $J_{X,Y}$ induce
off-diagonal mixing with states with $\Omega \neq 1$, which are
far in energy for this Hund's case (c) molecular state. We can
thus write the energy shift due to hfs, correct to second order in
$H_{\mathrm{hfs}} \approx A_{\parallel}J_{\mathrm{e}Z}I_{Z}$, as
\begin{align}
\Delta E_{\mathrm{hfs}}&\approx
\left<FIJM_{F}\Omega|A_{\parallel}J_{\mathrm{e}Z}I_Z|FIJM_{F}\Omega\right>
\nonumber \\
&\phantom{\approx}+ \sum_{J'=J\pm 1}\frac{
|\left<FIJ'M_{F}\Omega|A_{\parallel}J_{\mathrm{e}Z}I_Z|FIJM_{F}\Omega\right>|^{2}
}{E_{FIJ}-E_{FIJ'}}. \label{a1}
\end{align}
The matrix elements are determined from:
\begin{align}
&\phantom{=}
\left<FIJ'M_{F}\Omega|J_{eZ}I_{Z}|FIJM_{F}\Omega\right> \nonumber\\
&=(-1)^{F+I+J} \left\{\begin{matrix}
F & I & J' \\
1 & J & I
\end{matrix}\right\}
\sqrt{I(I+1)(2I+1)} \nonumber \\
&\phantom{=}\times (-1)^{J'-\Omega} \sqrt{(2J'+1)(2J+1)}
\left(\begin{matrix}
J' & 1 & J\\
-\Omega & 0 & \Omega
\end{matrix}\right)
\Omega.
\end{align}
Eqn. \ref{a1} then simplifies to:
\begin{align}
\Delta E_{\mathrm{hfs}}^{F=J\pm 1/2}= &\pm
\frac{A_{\parallel}}{2(F+1/2)} \nonumber \\
&\mp
\frac{A_{\parallel}^{2}}{8B_{v=5}}\frac{(F-1/2)(F+3/2)}{(F+1/2)^{3}}.
\label{eqn:hfs1}
\end{align}

Initial assignments of each line were possible using estimates of
the $B(1)$ state energies by extrapolation from the data of Ref.
\cite{martin}, and taking into account the known natural
abundances of each Pb isotope.  The line splitting due to hfs in
$^{207}$PbO is clearly visible in two transitions, corresponding
to the $Q1$ and $Q2$ lines, in Fig.{\ref{fig:scan1}}. The
unperturbed line positions of these hfs-split lines could be
located initially by extrapolating from $^{207}$PbO lines with
higher values of $J$, which exhibit much smaller hfs splitting.
The sign of $A_{\parallel}$ is found to be positive by visual
inspection of Fig. {\ref{fig:scan1}}, using the following
observations: the largest peak, which is closest to the
unperturbed line center, must arise from the $F=J+1/2$ state (see
Eqn.\ref{eqn:hfs1}); and in addition, if $A_{\parallel}>0$ ($<0$),
then this line should lie higher (lower) in energy than the
unperturbed line center.

Once the lines were assigned, each scan was analyzed with a
multiparameter fit in which the modulated absorption lines were
approximated as as sum over second derivatives of a gaussian, with
a separate gaussian for each value of $m_J$ in the excited state.
With the isotopic abundances fixed, several fit parameters were
used to characterize the height and the width of the various
transitions; these included the ratio of $P$:$Q$ and $R$:$Q$
transition strengths, the value of $\mu_{B(1)}$, the Doppler
width, and a measure of the electric field inhomogeneity. To
extract line positions, the energy $E^{Z,\mathrm{A}}(v,J)$ of
electronic state $Z$ ($Z=X,B$) with vibrational and rotational
quantum numbers $v$ and $J$, respectively, for the
$^\mathrm{A}$PbO isotopic species was written as
\begin{align}
E^{Z,\mathrm{A}}(v,J)= &G^{Z,\mathrm{A}}_{v}+ B^{Z,\mathrm{A}}_{v}
J\left(J+1 \right) - D^{Z,\mathrm{A}}_{v}J^2\left(J+1\right)^2 \nonumber \\
 &+ E^{X,\mathrm{A}}(v=0,J=0).
\end{align}
The energies of the $X$ state sublevels were taken from the
precise data of Ref. \cite{martin}. The $B$ state values were
determined by fitting, with separate values of the rotational
constants used for the $\Omega$-doublet levels with $e$ and $f$
character.  For $^{207}$PbO, the hfs constant $A_{\parallel}$ was
included as an additional fit parameter.

As described above, determination of $A_{\parallel}$ was the
primary goal of our work.  Fitting the data set yielded
$A_{\parallel} = 5.01(7)$ GHz, where the dominant uncertainty is
from the spread in the fit results from different scans, with
smaller contributions from variations in the fit results from
changes in the laser power correction and other input parameters.
This value is in good agreement with the {\it ab initio}
prediction in \cite{titov}.  We also find
$G^{B,208}_{v=5}=24600.085(8)$ cm$^{-1}$, (where the uncertainty
is from the wavemeter absolute calibration specification,
which we did not verify independently,
and from the extraction of the line centers); $B^{B,208}_{v=5}=0.24939(20)$ cm$^{-1}$;
and $\mu_{B(1)} = 4(1)$ D.  Our data is insensitive to the small
parameters $D^{B,208}_{v=5}$ and $q$.

The shifts between the values of $G^{B,\mathrm{A}}_{v=5}$ for the
different isotopes arise from changes in both the rovibrational
and electronic energies.  In order to separate these effects, it
is necessary to evaluate the absolute ro-vibrational energies of
the $B(1)(v=5)$ and the $X(0)(v=0)$ states.  We accomplish this
using a Dunham expansion :
\begin{equation}
E^{Z,\mathrm{A}}(v,J)=
T^{Z,\mathrm{A}}_{e}+\sum_{j,k}Y^{Z,\mathrm{A}}_{jk}
\left(v+1/2\right)^{j} \left( J \left(J+1 \right) \right)^{k}.
\label{eqn:dunham}
\end{equation}
Here the isotopic changes in rovibrational energy are taken into
account by writing $Y^{Z,\mathrm{A}}_{jk} =
Y^{Z,208}_{jk}\rho_{\mathrm{A}}^{-(j/2 + k)}$, where
$\rho_{\mathrm{A}} \equiv \mu(\mathrm{A}) / \mu(208)$ and
$\mu(\mathrm{A})$ is the molecular reduced mass of
$^\mathrm{A}$PbO. The coefficients in the Dunham expansion are
simply related to the previously described parameters.  For
example, $G^{B,\mathrm{A}}_{v=5}
=T^{B,\mathrm{A}}_{e}+\sum_{j}Y^{B,\mathrm{A}}_{j0}
\left(5+1/2\right)^{j}-\sum_{j}Y^{X,\mathrm{A}}_{j0}
\left(0+1/2\right)^{j}$.

To evaluate the rovibronic part of the isotopic shifts, we take
the values of $Y^{X,208}_{jk}$ from Ref.\cite{martin}. The Dunham
coefficients $Y^{B,208}_{jk}$ given in Table {\ref{tab1}} were
evaluated using the data of Ref. \cite{martin} on the three lowest
vibrational levels $(v=0,1,2)$ of the $B(1)$ state, in combination
with our determination of $G^{B,\mathrm{208}}_{v=5}$ and
$B^{B,\mathrm{208}}_{v=5}$.  Although inclusion of our data shifts
the values of individual Dunham coefficients noticeably (as
compared to their values using only the data from Ref.
\cite{martin}), the resulting change in the ro-vibronic part of
the isotope shifts is small: $\lesssim 100$ MHz, comparable to the
uncertainty in our determination of individual line positions.   
Due to our limited range in $J$, we have negligible sensitivity to
isotopic differences in the rotational constants.

\begin{table}
\caption{\label{tab1} Dunham coefficients $Y^{B,208}_{ij}$, in
cm$^{-1}$.  The value given in the place for $Y_{00}$ is actually
$T_e + Y_{00}$. }
\begin{ruledtabular}
\begin{tabular}{c r r}
i~~$\backslash$~~j & 0 & 1\\ \hline
0 &  22282.4297(60) & 0.26465(1) \\
1 & 498.497(13)    & -0.002561(12) \\
2 & -2.331(63) & -3.4(4)$\times 10^{-5}$ \\
3 & 0.03638(80) &  \\
\end{tabular}
\end{ruledtabular}
\end{table}

With this determination of the rovibronic part of the isotope
shift, it is possible to extract the isotopic shifts in the
electronic energies.  Defining $\Delta \nu
(\mathrm{A}-\mathrm{A}') \equiv \left(
T^{B,\mathrm{A}}_{e}-T^{X,\mathrm{A}}_{e} \right) - \left(
T^{B,\mathrm{A}'}_{e}-T^{X,\mathrm{A}'}_{e} \right)$, we find
$\Delta \nu (208-207) = -270(50)$ MHz and $\Delta \nu (208-206) =
-720(60)$ MHz. 
In a second approach,
line centers from two different isotopes were extracted
with no constraints on their relative positions or amplitudes. Subtracting 
off the rovibronic part of the separation expected between the lines from 
Eqn.{\ref{eqn:dunham}} gave consistent results for the electronic 
part of the isotope shifts. To account for these electronic shifts, we write
\begin{equation}
T^{Z,\mathrm{A}}_{e}= T^{Z,P}_{e}\left(1 + V^{Z}\left\langle
r_{\mathrm{A}}^2 \right\rangle +
\Delta^{Z}\frac{m_e}{M_\mathrm{A}} \right).
\end{equation}
In this expression, $T^{Z,P}_{e}$ is the electronic energy of
state $Z$ for a hypothetical point-like nucleus of infinite mass,
and the parameters $V^{Z}$ and $\Delta^{Z}$ characterize the
field-shift and mass-shift parts, respectively, of the electronic
isotope shift \cite{knockel,king}; here $\left\langle
r_{\mathrm{A}}^2 \right\rangle$ and $M_{\mathrm{A}}$ are the
mean-square charge radius and the mass, respectively, of the
$^{\mathrm{A}}$Pb nucleus, and $m_e$ is the electron mass.  We use
the known difference in mean-square charge radii between the
different Pb isotopes \cite{thompson} to separate our measured
values for the electronic shift into field- and mass-shift
contributions.  We find $V^B - V^X = 1.0(7)\times 10^{-5} $
fm$^{-2}$, and $\Delta^B - \Delta^X = 8.5(3.1) \times 10^{-6}$.
The normal part of the mass shift (due to the change in the
electron reduced mass) is negligibly small; hence the observed
shift is a specific mass shift, due to electron correlations. This
shift is unusually large, although similar shifts are known to
occur in complex atoms \cite{king}.

In summary, we have measured some detailed spectroscopic
parameters of the $X(0)(v=0) \rightarrow B(1)(v=5)$ transition of
PbO.  The measured value of the $B(1)$ state hfs provides an
important data point for checking calculations of the effective
electric field acting on an electron EDM in excited states of PbO.

We thank M. Kozlov for useful conversations. This work was
supported by NSF Grant PHY0244927, and the David and Lucile
Packard Foundation.

%

%

\begin{thebibliography}{99}

\bibitem{misha}
M.G. Kozlov and D. DeMille, Phys. Rev. Lett. {\bf{89}}, 133001
(2002).
\bibitem{titov}
T.A. Isaev, A.N. Petrov, N.S. Mosyagin, A.V. Titov, E. Eliav, and
U. Kaldor, Phys. Rev. A {\bf{69}}, 030501(R) (2004); A.N. Petrov,
A.V. Titov, T.A. Isaev, N.S. Mosyagin, and D. DeMille, Phys. Rev.
A {\bf{72}}, 022505 (2005).
\bibitem{regan}
B.C. Regan, E.D. Commins, C.J. Schmidt, and D. DeMille, Phys. Rev.
Lett. {\bf{88}}, 071805 (200).
\bibitem{daved} D. DeMille, F. Bay, S. Bickman, D. Kawall, D. Krause, Jr., S.E.
Maxwell, and L.R. Hunter, Phys. Rev A {\bf{61}}, 052507 (2000); D.
Kawall, F. Bay, S. Bickman, Y. Jiang, and D. DeMille, Phys. Rev.
Lett. {\bf{92}}, 133007 (2004).
\bibitem{larry}L.R. Hunter, S.E. Maxwell, K.A. Ulmer, N.D. Charney, S.K. Peck,
D. Krause, Jr., S. Ter-Avetisyan, and D. DeMille, Phys. Rev. A
{\bf{65}}, 030501(R) (2002).
\bibitem{egorov} D. Egorov, J.D. Weinstein, D. Patterson, B. Friedrich, and J.M.
Doyle, Phys. Rev. A {\bf{63}}, 030501(R) (2001).
\bibitem{malcolm} A.S. Arnold, J.S. Wilson, and M.G. Boshier, Rev. Sci. Instrum.
{\bf{69}}, 1236 (1998); C.J. Hawthorn, K.P. Weber, and R.E.
Scholtena, Rev. Sci. Instrum. {\bf{72}}, 4477 (2001).
\bibitem{watanabe}T. Watanabe, and Y. Amako, J. Chem. Phys. {\bf{106}}, 3891 (1997).
\bibitem{martin}
F. Martin, R. Bacis, J. Verg\`es, J. Bachar, and S. Rosenwaks,
Spectrochim. Acta {\bf{44A}}, 889 (1988).
\bibitem{howell}
H. G. Howell, Proc. R. Soc. London A {\bf{153}}, 683 (1936); E. A.
Dorko {\it{et al.}}, Chem. Phys. {\bf{102}}, 349 (1986).
\bibitem{watson}
J. K. G. Watson, J. Mol. Spectrosc. {\bf{80}}, 411 (1980).
\bibitem{schlembach}
J. Schlembach and E. Tiemann, Chem. Phys. {\bf{68}}, 21 (1982).
\bibitem{knockel}
H. Kn\"ockel, U. Lindner, and E. Tiemann, Mol. Phys. {\bf{61}},
351 (1987).
\bibitem{king} W.H. King, {\it{Isotope Shifts in Atomic Spectra}}
(Plenum Press, New York, 1984).
\bibitem{thompson}
R.C. Thompson \textit {et al.}, J. Phys. G {\bf{9}}, 443 (1983).
\end{thebibliography}
\end{document}